# DATA ACQUISITION SYSTEM WITH SHARED MEMORY NETWORK

T. Masuda, T. Fukui, R. Tanaka, A. Yamashita
SPring-8, Hyogo 679-5198, Japan

*Abstract*

The SPring-8 control system provides data acquisition software called Poller/Collector for periodic data taking. Its polling scheme does not support taking data synchronized with an event, and simultaneity in the data from the different VME systems is not guaranteed. Recently, requirement for fast and event-driven data acquisition has increased, and the synchronization of a set of data from the different VME systems is requested. We added new data-taking software to the standard framework with a shared memory network. We designed the software on the VME by updating the original framework that was already used for feedback control. The data filling process on a PC stores sets of the synchronized data on a shared memory board to the database. The shared memory boards provide 250Mbit/s transmission rate and are linked with optical fiber cables. We will install the system to the linac beam position monitors data acquisition as the first application.

## 1 INTRODUCTION

In the SPring-8 control framework, Poller/Collector software has been used for periodic data acquisition [1]. Originally, Poller/Collector was developed to collect the storage ring equipment data, that is, its scheme was designed on the assumption that the target data was almost constant. Poller/Collector was extended as a data acquisition system for the beamline control, the injector linac and synchrotron control afterwards. Poller processes on the VME CPUs periodically acquire a set of the specified data. The Pollers store the acquired set to ring buffers on the memory of the CPU board. The Collector Client collects only the newest data set from the ring buffers. In this scheme, it is not guaranteed to collect a set of data that satisfies simultaneity between the different VME systems.

Recently, we need to acquire the synchronized data with an event from the distributed VMEs as fast as possible, for example, data of beam position monitors (BPMs) of the linac. We added new data acquisition software to the standard framework with a shared memory network. The network can share the memory image with no software overhead. By introducing this network, we can expect that the network can provide a real-time communication among some processes running on the different computers.

## 2 SHARED MEMORY NETWORK

### 2.1 Features

To build a shared memory network, we introduced the products provided by Advanet Inc.[1], that is, Advme1522A boards for VMEbus (as shown in Figure 1), Adpci1523A boards for PCI bus and E-045A HUBs for the shared memory network. The boards are connected with 250Mbit/s on multi-mode optical fiber cables. An E-045A HUB enables the shared memory network to connect in a star topology. The shared memory board can reflect the memory image even if the bus type of the board is different. The maximum size of a shared memory is 8MB. The board can generate an interrupt to the specified board or to all the boards at the same time on the network.

Device drivers were developed for the shared memory boards. We can control the boards and read/write data from/to the memory on the board through the device drivers with an *ioctl* system call. We prepared Solaris and HP-RT device drivers for the Advme1522A, and Solaris and Linux ones for the Adpci1523A. We developed application program interface (API) functions for programmers to hide the system differences. We define a protocol for the applications that use the shared memory network.

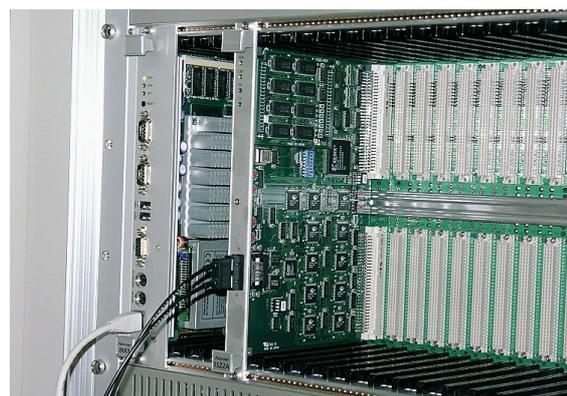

Figure 1: Picture of an Advme1522A shared memory board with an Advme8001 CPU board.

### 2.2 Performance

We measured effective transmission speed of the shared memory network. The following test systems were used for the measurements:

---

[1] http://www.advanet.co.jp

- DELL PC Optiplex G1 (333MHz Celeron) + Linux2.2.16 + Adpci1523A.
- HP9000/743rt CPU board (64MHz PA-7100LC) + HP-RT 2.21 + Advme1522A.
- Advme8001 CPU board (600MHz Pentium III) + Solaris 7 + Advme1522A.

For the evaluation, two shared memory boards were installed in the same bus of a computer and connected to each other through an E-045A HUB. We evaluated the transfer rate as a function of the data size by measuring the difference between the start time of writing to a board and the end time of its reflection on the other board.

Figure 2 shows the result of the PC system. The transfer speed is saturated at 3.4MB/sec. The saturated transfer rates of all the test systems are listed in Table 1. According to the coincident results of the PC and the HP9000/743rt systems, the effective transfer rate of the shared memory network can be estimated to be 3.40MB/s. In the system of Advme8001, an overhead due to a PCI-to-VME bus-bridge may cause a drop in speed.

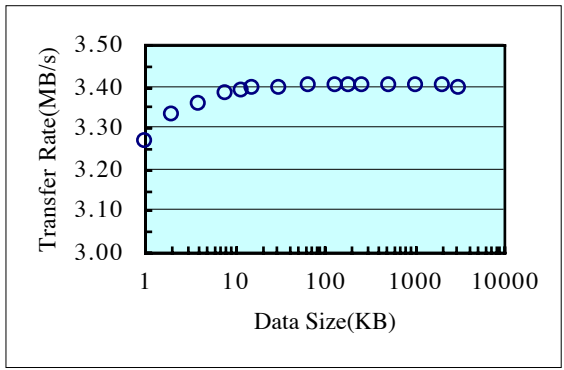

Figure 2: Measurement of the transfer rate on the DELL PC system.

Table 1: Saturated transfer rates

| System | Transfer Rate |
| --- | --- |
| DELL PC Optiplex G1 + Linux2.2.16 + Adpci1523A | 3.40MB/s |
| HP9000/743rt + HP-RT2.21 + Advme1522A | 3.40MB/s |
| Advme8001 + Solaris 7 + Advme1522A | 3.18MB/s |

# 3 SOFTWARE

## 3.1 Software Structure

A schematic diagram of the new data acquisition software is shown in Figure 3. We designed the new software based on the original framework named the Equipment Manager Agent (EMA) [2]. It was already used for medium-speed feedback control. The scheme consists of event-driven EMAs (EMA-EVs), Event Generators (EVGENs) and a Filler. The EVGEN watches a designated event continuously and sends an event message to the EMA-EV whenever the event occurs. The EMA-EV is an event-driven data acquisition process created from the Equipment Manager (EM) that is used for the interactive control of equipment [3]. The EM controls an EMA-EV as a pseudo device.

The EMA-EV acquires a set of specified data immediately the event is notified by an EVGEN and puts it to a shared memory board. The EVGEN and the EM communicate with the EMA-EV by using an IPC message through the Message Server (MS) [4]. The filler process collects all the data on the shared memory network written by the EMA-EVs then commits them to a database.

## 3.2 EMA-EV

When the EMA-EV starts, it creates an EVGEN process then waits for the message. The message recognized by the EMA-EV is a "control" message from the EM and an "event" message from the EVGEN. By receiving the "control" message, the EMA-EV changes the state of the EVGEN to running or pausing. On the other hand, when the EMA-EV gets the "event" message, the EMA-EV starts data acquisition.

## 3.3 EVGEN

The EVGEN can be controlled only from the EMA-EV with System V signals. When the EVGEN observes an event, it sends an "event" message to the EMA-EV through the MS.

In the EVGEN framework, an application programmer can create an add-in function to impose an event source. An arbitrary event source is available to handle not only a hardware event but also a software event.

## 3.4 Filler

The "data filler" process running on a PC fetches data from the shared memory board and records them in a logging database periodically. The database keeps the data for further off-line analysis.

The filler can control the data acquisition system by sending "control" messages. In addition to the "start", "stop" and "destroy" command, the filler can pass a "create" command to instruct the EM to create the EMA-EV.

## 3.5 Project

A number of the independent data acquisition systems are operational on the same network at the same time. One data acquisition system is called as a "project". Each project has a unique name to distinguish it from others. The information about the project such as a filler ID, an EMA-EV ID, signal lists are stored in a database.

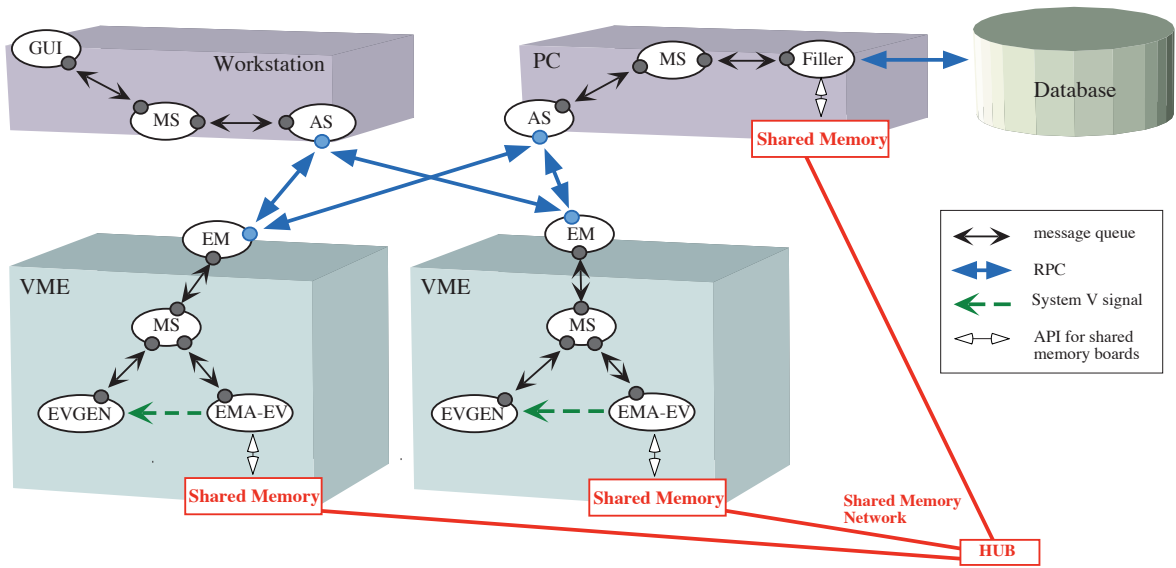

Figure 3: Schematic diagram of the new data acquisition software.

To share the project information among all the EMA-EVs and the filler, a file identified by a project name is generated from the database. When the filler and the EMA-EVs start, they read the target project file.

## 4 APPLICATION AND PLAN

We will apply the new data acquisition system to the BPM data acquisition for the injector linac at first. Currently, 31 BPMs are installed in the linac and beam transport lines. We are going to acquire the data of all the BPMs synchronized with one shot of electron beam. The six VME systems are prepared for the BPM data acquisition, and a PC is used for the filler process. Each VME system has one Advme1522A board controlled by an Advme8001 CPU board with Solaris 7. The PC operated by Linux has one Adpci1523A board. The seven shared memory boards are linked by optical fibers in a star topology. At present, we collect the data with 1Hz repetition of the linac operation.

We are planning to achieve 60Hz data acquisition cycle to meet the 60pps operation of the linac.

## 5 SUMMARY

We developed new software for the synchronized data acquisition with an event. The new data acquisition software was developed based on the existing software framework of the EMA. To achieve high throughput, the shared memory network was introduced in the data acquisition system. The effective transfer rate of the shared memory network concludes around 3.4MB/s.

We finished the development, and the test system worked well. We plan to apply this new data acquisition system to the BPM data collection of the injector linac with 1Hz. The final target of the repetition is about 60Hz.

## REFERENCES

[1] A. Taketani *et al.*, "Data acquisition system with database at the SPring-8 storage ring", ICALEPCS'97, Beijing, China, 1997, p.437.
[2] A. Taketani *et al.*, "Medium-speed feedback software based on the existing control system", ICALEPCS'97, Beijing, China, 1997, p.486.
[3] A. Taketani *et al.*, "Equipment Manager of the VME Control System for the SPring-8 Storage Ring", ICALEPCS'95, Chicago, USA, 1995, p.625.
[4] R. Tanaka *et al.*, "The first operation of control system at the SPring-8 storage ring", ICALEPCS'97, Beijing, China, 1997, p.1.